\documentclass[showpacs,preprintnumbers,amsmath,nofootinbib,aps]{revtex4}

\usepackage{graphicx,epsf, epsfig, amssymb}
\usepackage{bm}
                                                                                                                           
\begin{document}
                                                                                                                           
\title{QCD and spin effects in black hole airshowers}

\author{Marco Cavagli\`a}
\email{cavaglia@olemiss.edu}
\affiliation{Department of Physics and Astronomy, University of Mississippi,
University, MS 38677-1848, USA}

\author{Arunava Roy}
\email{arunav@olemiss.edu}
\affiliation{Department of Physics and Astronomy, University of Mississippi,
University, MS 38677-1848, USA}
                                                                                                                           
\date{\today}

\begin{abstract}
In models with large extra dimensions, black holes may be produced in high-energy particle
collisions. We revisit the physics of black hole formation in extensive airshowers from
ultrahigh-energy cosmic rays, focusing on collisional QCD and black hole emissivity effects.
New results for rotating black holes are presented. Monte Carlo simulations show that QCD
effects and black hole spin produce no observable signatures in airshowers. These results
further confirm that the main characteristics of black hole-induced airshowers do not depend
on the fine details of micro black hole models.
\end{abstract}
\maketitle
                                                                                                                           
\section{Introduction\label{intro}}

In an effort to unify gravity with the other fundamental forces, we are faced with the
hierarchy problem. The fundamental scale of gravity is about 17 orders of magnitude higher
than the TeV scale, where electromagnetic and weak forces unify. The hierarchy problem may be
solved by the introduction of large extra dimensions \cite{Arkani-Hamed:1998rs}. In this
model, the Planck scale $M_{\rm Pl}$ is related to the fundamental scale of gravity
$M_\star\sim 1$ TeV by the relation $M_{\rm Pl}^2\sim V_n M_{\star}^{n+2}$, where $V_n$ is
the volume of the extra $n$-dimensional space. Gravity is a strong force in the
higher-dimensional spacetime but appears weak to a four-dimensional observer due to its
``leakage'' in the extra dimensions. Gravitons may propagate in all dimensions (bulk).
Compatibility with known sub-TeV physics restricts the propagation of all Standard Model (SM)
fields to three spatial dimensions (brane).

One of the effects of the increased strength of gravity would be the production of TeV-scale
Black Holes (BHs) in high-energy particle collisions \cite{Banks:1999gd}. Micro BHs could be
produced in man-made particle colliders, e.g.\ the Large Hadron Collider
\cite{Giddings:2001bu,Dimopoulos:2001hw, Cavaglia:2006uk}, or naturally in Earth's atmosphere
by Ultra High Energy Cosmic Rays (UHECRs) interacting with air nucleons
\cite{Feng:2001ib,Anchordoqui:2001cg,Ahn:2003qn,Cafarella:2004hg}. (For reviews, see Refs.\
\cite{Cavaglia:2002si}.)  Once formed, these BHs would immediately decay through loss of
excess multipole moments (balding phase), Hawking emission \cite{Hawking:1974sw} (evaporation
phase) and final $n$-body decay or remnant production (Planck phase). SM fields may be
originated in each of these stages, providing a means to detect the BHs. For atmospheric
events, the visible imprint would be an extensive airshower initiated by these SM quanta.

The characteristics of BH-induced airshowers can be investigated with Monte Carlo techniques.
A Monte Carlo code for BH formation and airshower generation is GROKE   \cite{Ahn:2005bi}.
GROKE simulation of BH events proceeds in three stages. First, the BH is formed by the
collision of an UHE neutrino and an air nucleon parton. Some of the Center-of-Mass (CM)
energy is lost in the process (about 40\% for head on collisions, monotonically increasing
with the impact parameter \cite{Yoshino:2002tx,Yoshino:2005hi}). SM unstable particles from
the BH decay and the nucleon remnant are hadronized using a high-energy physics program for
event generation (PYTHIA) \cite{Sjostrand:2006za}. PYTHIA's output is then injected into a
simulator of extensive airshowers (AIRES) \cite{Sciutto:1999rr}. Simulations show that BH
airshowers generally rise faster, have broader peak and higher variation in the total energy 
than SM airshowers because of the ``democratic'' nature of BH decay. BH events are also
characterized by a larger muonic content at the ground compared to SM events due to the
dominant hadronic channel in the BH evaporation phase. A complete discussion of BH signatures
can be found in Refs.\ \cite{Ahn:2005bi,Ahn:2003qn}.

BH searches require the identification of observational signatures that do not depend on the
fine details of the model. This can be achieved by improving the theoretical description of
the event and then testing the stability of the airshower characteristics against these
theoretical refinements. Previous investigations \cite{Ahn:2005bi,Ahn:2003qn} neglected or
approximated various aspects of the physics of BH formation and decay such as QCD effects, BH
spin and particle emissivities. QCD effects in the fragmentation process may lead to changes
in the amount of visible energy deposited in the airshower by the nucleon remnant. Changes in
particle emissivities due to spacetime dimensionality or BH rotation may affect rapidity,
peak variation and muon content of airshowers. The aim of this paper is to check the
stability of BH airshowers characteristics when these effects are included. The result of our
investigation is that inclusion of collisional QCD effects, changes in particle emissivity
and BH rotation do not significantly affect the BH airshower development: Observational
signatures of BH events are robust. Natural units are used throughout the paper with
$M_{*}=1$.

\section{QCD and emissivity effects\label{QCD}}

QCD effects in airshower generation and development include initial- and final-state
radiation, fragmentation, and the hadronization process of nucleon remnant and unstable
quanta. Since PYTHIA is designed to handle initial- and final-state radiation, multiple
scattering, beam remnant and hadronization, these effects can be investigated by modifying
the GROKE code \cite{Ahn:2005bi} to include the BH event as a PYTHIA external process. 

The BH airshower is initiated by the decay products of the BH, the nucleon remnant and jets
from initial- and final-state radiation. The colliding parton is taken from the Parton
Distribution Functions at a very high energy scale ($\gtrsim$ TeV). This implies that it
typically emits quite hard initial-state radiation before it collides to form a black hole,
resulting in additional jets. In addition, the nucleon remnant undergoes soft and semi-hard
multiple scatterings which will contribute to the airshower. Previous investigations of BH
airshowers did also not take into account color conservation. However, the nucleon remnant is
color connected to the BH decay products; the color flow will also hadronize into more jets.
In our investigation, color flow is implemented in the $N_C \to \infty$ limit of QCD
\cite{Sjostrand:2006za}. 

Total multiplicity ($N$) and multiplicity per species ($N_i$) of the BH decay are essential
to determine the airshower characteristics. Earlier studies \cite{Ahn:2005bi,Ahn:2003qn} used
approximated thermally-averaged emissivities (graybody factors) for the evaporation phase.
Recently, exact graybody factors for higher-dimensional nonrotating BHs were calculated in
Ref.\ \cite{Cardoso:2005mh}. Changes in greybody factors are specially relevant for 
higher-dimensional spacetimes, where graviton emission is highly enhanced. These results  are
implemented in GROKE following Ref.\ \cite{Cavaglia:2006uk}. The total multiplicity in the
evaporation phase is
\begin{equation}
N=\frac{(n+1)S}{4\pi}\,\frac{\sum_i c_i{\cal R}_i\Gamma_{{\cal R}_i}} {\sum_j c_j{\cal
P}_j\Gamma_{{\cal P}_j}}\,,
\label{multin}
\end{equation}
where $S$ is the initial entropy of the BH, $c_{i}$ are the degrees of freedom of the $i$-th
species, and $\Gamma_{{\cal P}_i}$ and $\Gamma_{{\cal R}_i}$ are the fraction of radiated
power and the emission rate per degree of freedom, respectively. The decay multiplicity per
species is
\begin{equation}
N_i=N\,\frac{c_i{\cal R}_i\Gamma_{{\cal R}_i}} {\sum_j
c_j{\cal R}_j\Gamma_{{\cal R}_j}}.
\label{multii}
\end{equation}
The use of exact greybody factors (nonrotating case) leads to a slight reduction in the
output of visible energy and an enhancement of graviton multiplicity in the evaporation phase
compared to previous studies. These effects are generally of order $\sim 1$ or less.

\begin{figure*}
\centerline{\null\hfill
    \includegraphics*[width=0.4\textwidth]{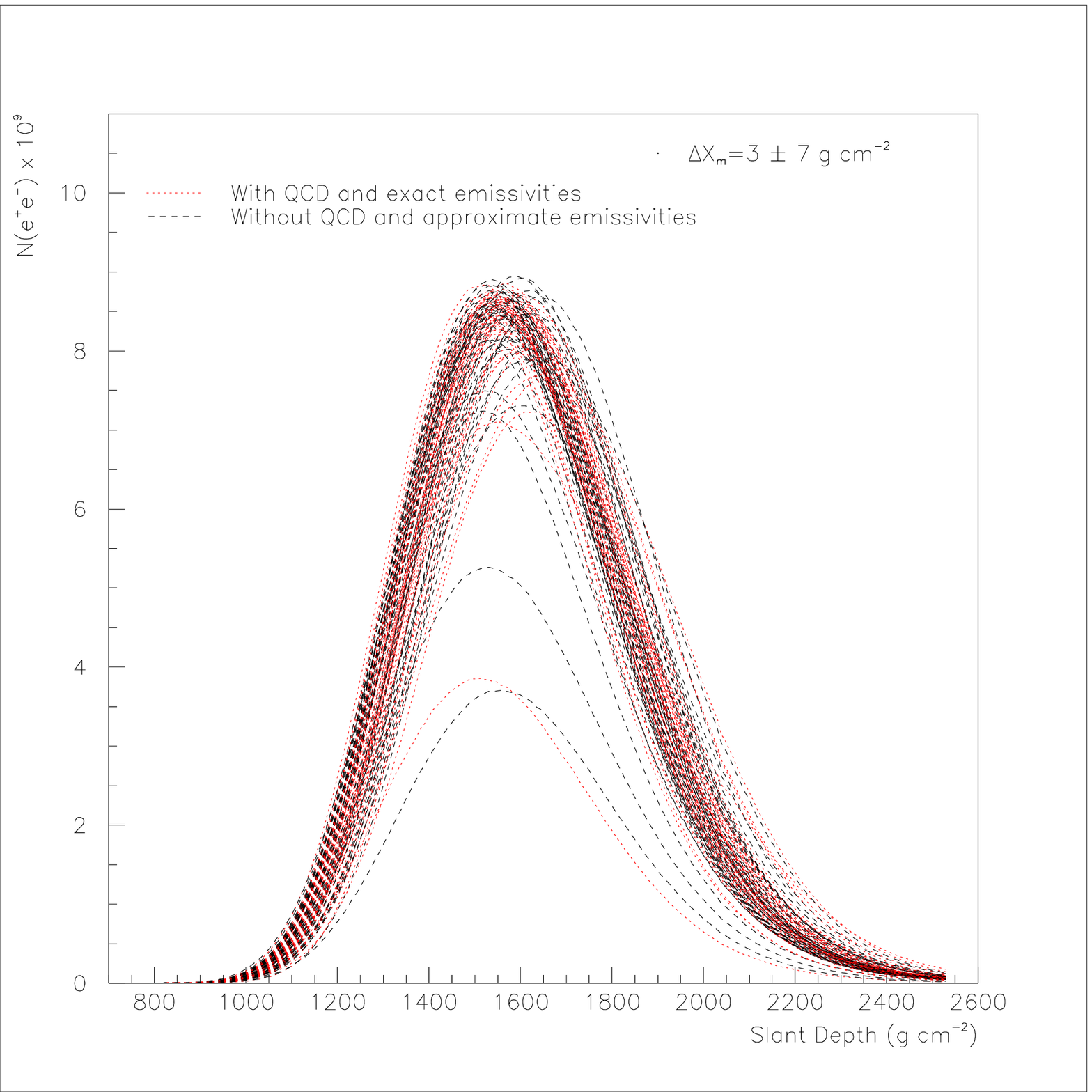}
    \hfill
    \includegraphics*[width=0.4\textwidth]{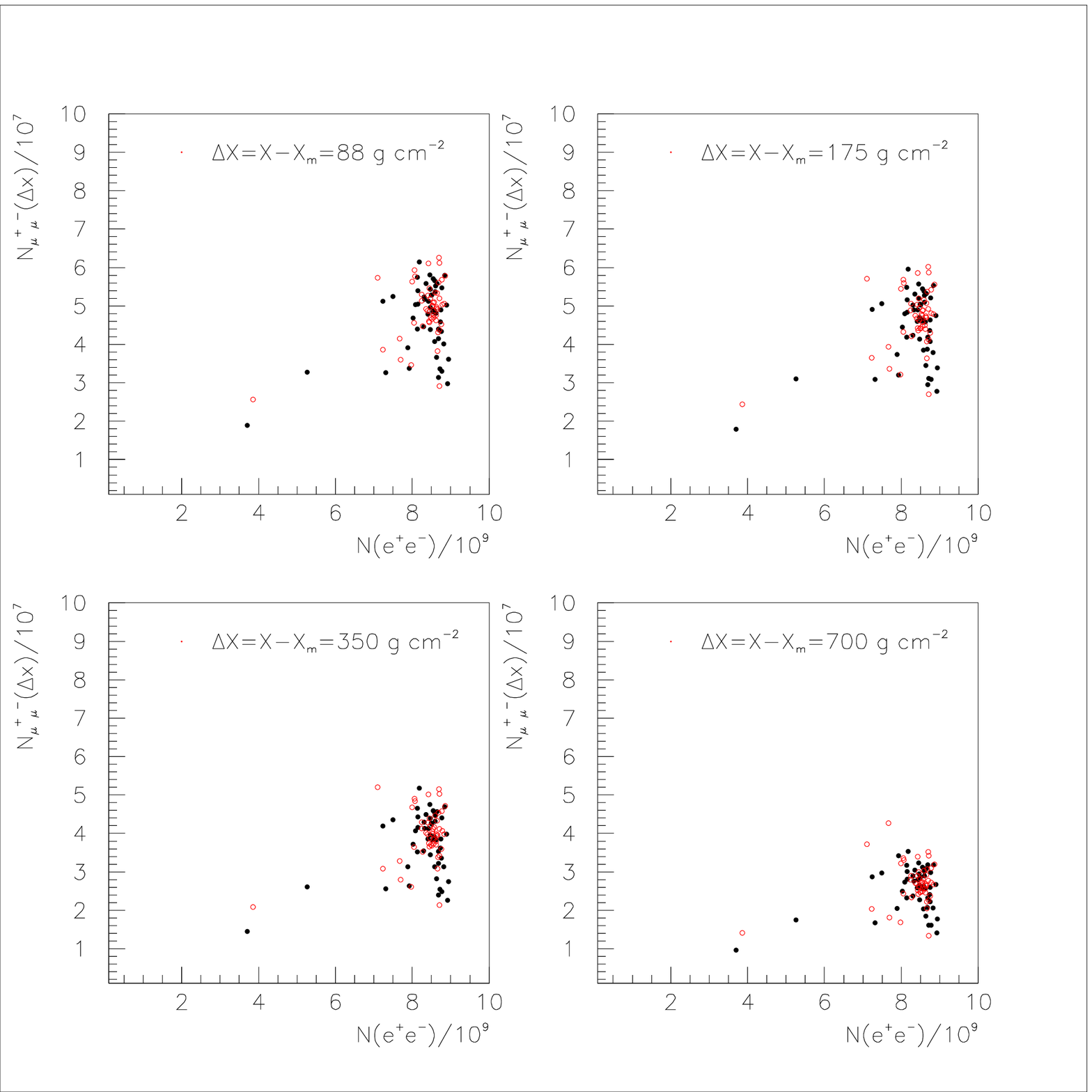}
    \hfill}
\caption{Left panel: Number of $e^+e^-$ pairs vs.\ slant depth for the longitudinal
development of 50 ten-dimensional BH airshowers. The neutrino primary energy is $E_\nu =
10^{19}$ TeV, the altitude of the first interaction depth is 10 km (slanth depth 780
g~cm$^{-2}$) and the zenith angle is 70${}^\circ$. BH airshowers without (with) QCD and
emissivity effects are shown by black dashed (red dotted) curves. Right panel: Number of $\mu^+\mu^-$
pairs at various atmospheric depths $X_m + \Delta X$ vs.\ the number of $e^+e^-$ at the
shower maximum. BH airshowers without (with) QCD and emissivity effects are shown by black filled
(red empty) circles. The observation depth increases from left to right and top to bottom.}
\label{FIG1}
\end{figure*}

QCD and emissivity effects on the BH airshower development can be determined by looking at
the longitudinal development of the $e^+e^-$ component of the airshower and the muonic
content at ground. The left panel of Fig.\ \ref{FIG1} compares 50 BH airshowers with and
without QCD and emissivity effects (primary neutrino energy $E_\nu=10^{19}$ eV, ten spacetime
dimensions). The average depth of the airshower maxima $X_m$ is not significantly affected by
the inclusion of initial- and final-state radiation, color conservation and exact emissivities. This can be qualitatively
explained by looking at the energy distribution of the BH 
airshower initiatiors after the hadronization and fragmentation process. PYTHIA's output
shows that the additional jets from initial- and final state radiation are generally too soft
to affect the airshower development, which is mainly determined from the evolution of the
nucleon remnant and the hard hadronic jets from the BH evaporation. The implementation of
color conservation slightly changes the details of the hadronization process of previous
studies. However, the main characteristics of the airshower depend on the hadronic nature of
the event rather than the details of the fragmentation. Differences in the fragmentation
model are washed out by uncertainties in the airshower development. A similar qualitative
explanation applies to the emissivity effects. Even for massive BH events, when quanta from
the evaporation phase dominate over quanta from the Planck phase and the nucleon remnant,
changes due to the use of exact greybody factors are too small to produce an observable
effect in the airshower profile. Identical conclusions are reached by comparing the number of
muons at various depths vs.\ the number of $e^+e^-$ pairs at the airshower maximum (right
panel of Fig.\ \ref{FIG1}). 

\section{Rotating black holes\label{spin}}

Up to now, simulations of BH airshowers have focused on Schwarzschild BHs
\cite{Ahn:2005bi,Ahn:2003qn,Cafarella:2004hg}. However, BHs created in collisions with
nonzero impact parameter are expected to be spinning.  Since the evaporation process depends
on the BH angular momentum, airshowers initiated by spinning BHs could be significantly
different from airshowers initiated by Schwarzschild BHs.  If the graviton emissivity for
rotating BHs is much higher than the emissivity of SM particles, only gravitons will be
emitted, making the BH undetectable. Although particle emissivities for higher-dimensional
rotating BHs are not fully known \cite{Duffy:2005ns}, the effects of rotation in the
evaporation phase can be estimated from results in four dimensions and for higher-dimensional
nonrotating BHs.

In four dimensions, the evaporation phase of a spinning BH with large angular momentum is
dominated by gravitons \cite{Page:1976}. The emissivity of spin-2 fields increases by a factor
$\sim 10^2 - 10^3$ more than the emissivity of lower spin particles when the angular momentum
increases from $J=0$ to the maximum value $J_{max}=M^2$, where $M$ is the mass of the BH. For
a random distribution of BH spins, the average increase in graviton emissivity is $\sim 10$
more than the other fields. Graviton emission also increases with the number of dimensions
due to a higher number of spin-2 helicity states. This has been shown quantitatively in Ref.\
\cite{Cardoso:2005mh} for nonrotating BHs. The graviton-to-SM emission ratio increases from
1:$10^3$ in four dimensions to 1:4 in eleven dimensions.

\begin{figure*}
\centerline{\null\hfill
   \includegraphics*[width=0.4\textwidth]{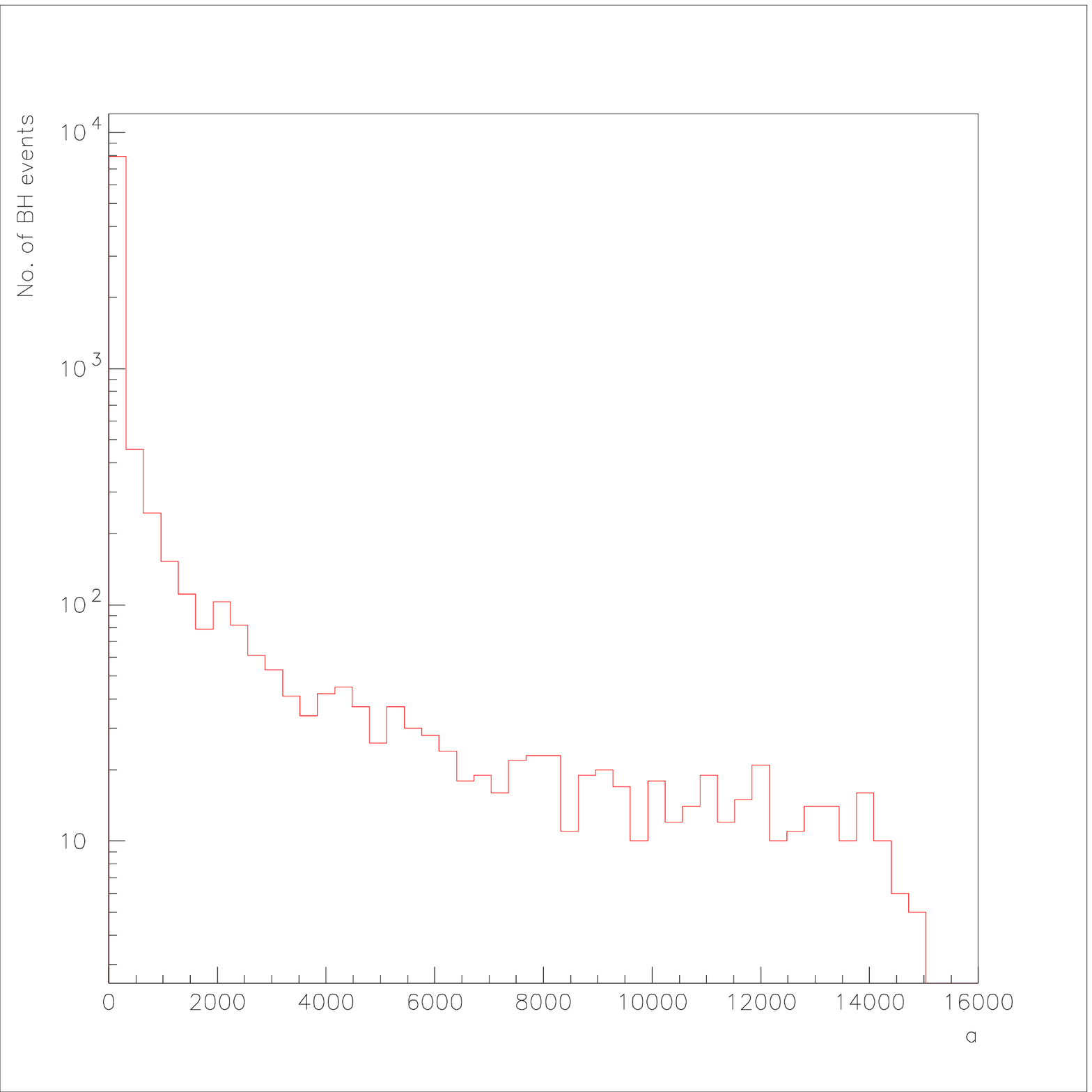}
   \hfill
   \includegraphics*[width=0.4\textwidth]{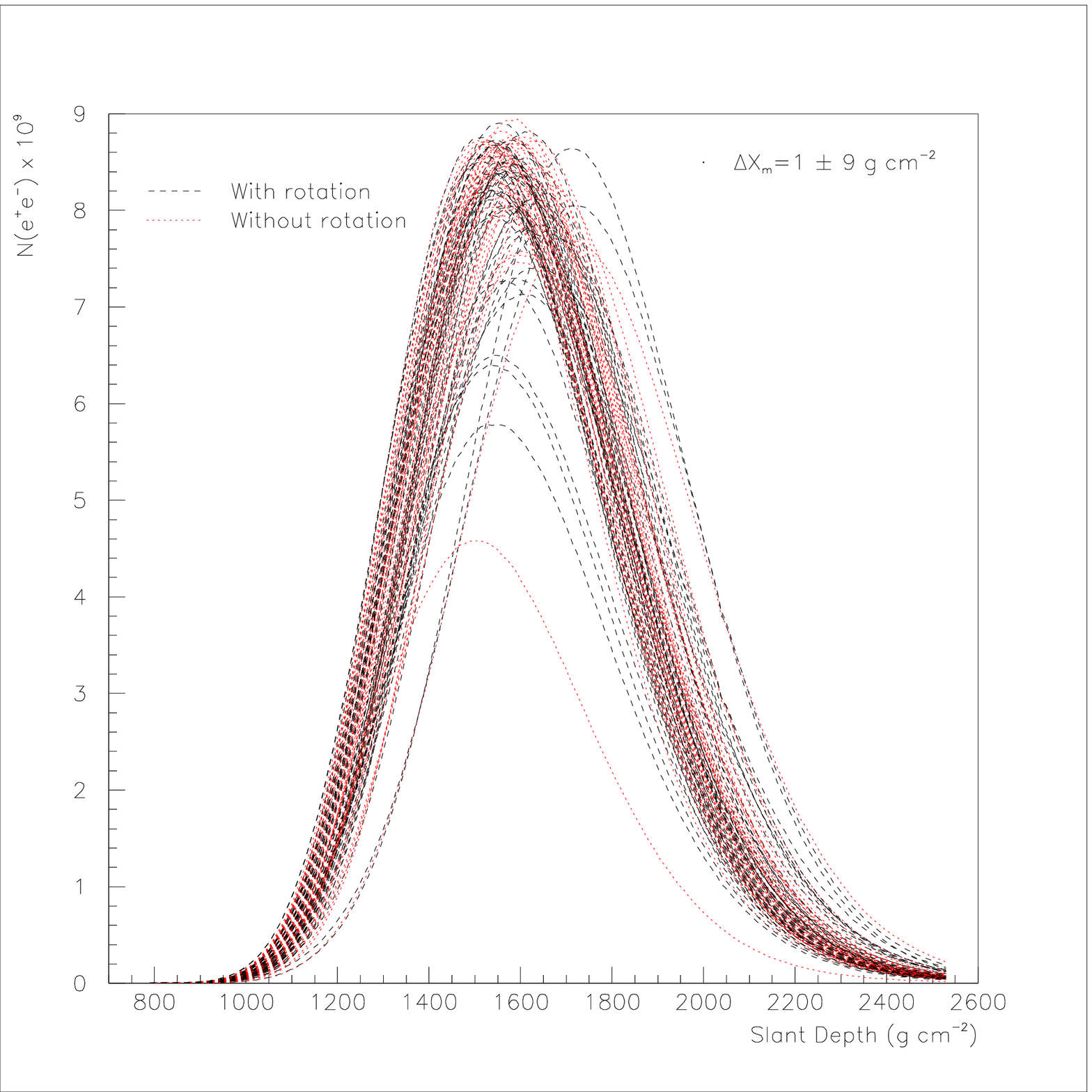}
   \hfill
  }
  \caption{Left Panel: Histogram of the number of events vs.\ $a$ for the decay of a ten-dimensional rotating BH.
  Right Panel: Shower profiles for both rotating and nonrotating BHs. The black dashed (red dotted) curves denote
  rotating (nonrotating) BH events.}
  \label{FIG2}
  \end{figure*}

The results above suggest a larger graviton greybody factor for higher-dimensional, spinning
BHs. The increase in graviton emissivity is specially relevant for ultra-spinning BHs. If
most of the BHs produced in UHECR collisions are low-spinning, graviton emission is likely to
increase on average by one order of magnitude more than the other fields. If most of the BHs
are ultra-spinning, graviton emission could be enhanced by several orders of magnitude. It is
thus crucial to determine the distribution of BH spins in airshowers. To this purpose, let us
define the parameter $a$ \cite{Giddings:2001bu}: 
\begin{equation}
a=\frac{D-2}{2}\frac{J}{MR}\,,
\label{ad}
\end{equation}
where $D$ is the number of dimensions and $R$ is the BH radius. The radius of the BH is related to the mass and angular momentum by the
relation \cite{Myers:1986un}
\begin{equation}
R=\frac{1}{\sqrt{\pi}} \left[\frac{8M}{1+a^2}\frac{\Gamma\left(\frac{D-1}{2}\right)}{D-2}
\right]^{1/(D-3)}.
\label{kerr}
\end{equation}
We consider the model of Ref.\ \cite{Yoshino:2005hi} for BH formation. The colliding
particles are described by boosted Schwarzchild solutions at fixed energy (Aichelburg-Sexl
shock waves) \cite{Aichelburg:1970dh}. The BH is formed when the two waves are superposed to
form a trapped surface. The mass of the BH is related to the CM energy of the colliding
particles, $E_{cm}$, by $M=E_{cm} y$, where $y$ depends on the impact parameter $b$ of the
collision. The ratio $J/M$ in Eq.\ (\ref{ad}) is
\begin{equation}
\frac{J}{M}=\frac{x r_0}{2 y},
\label{jm}
\end{equation}
where $x=b/r_0$ is the impact parameter normalized to $r_0=\left(4 \pi
E_{cm}/\Omega_{D-3}\right)^{1/(D-3)}$ and $\Omega_{D-3}$ is the area of the unit sphere in
$D-3$ dimensions. The parameter $a$ is the solution of the polynomial equation
\begin{equation}
a^{D-3}=C\frac{1+a^{2}}{y} \left( \frac{x}{y} \right)^{D-3}\,,
\label{aeq}
\end{equation}
where
\begin{equation}
C=\sqrt{\pi}\left(\frac{D-2}{4}\right)^{D-2}\frac{\Gamma(\frac{D-2}{2})}{\Gamma
(\frac{D-1}{2})}\,.
\end{equation}
The left panel of Fig.\ \ref{FIG2} gives the distribution of $a$ for 10,000 events in ten
dimensions. Most of the BH are formed with small angular momentum. This is somehow expected
because the BH production cross section is reduced by a factor of $(1+a^2)^{-2/(D-3)}$
compared to the nonrotating case \cite{Park:2001xc,Anchordoqui:2001cg}. According to the
discussion above, the increase in graviton emissivity relative to lower-spin fields can be
estimated to be about one order of magnitude compared the nonrotating case.

Simulations for spinning BHs with  $\Gamma_{{\cal R}_2}\sim\Gamma_{{\cal
P}_2}=10\times(\hbox{nonrotating}~\Gamma_{{\cal R}_2},\Gamma_{{\cal P}_2})$ show that most of
the emission is in the form of gravitons. However, the number of observable secondaries for
both rotating and nonrotating BH airshowers is stable and the difference in the profiles is
statistically not significant (right panel of Fig.\ 2). In a typical event, the bulk of the
collisional CM energy is carried by the nucleon remnant. Therefore, changes in particle
emissivities have generally minor effects on the airshower characteristics. Rare events
($\lesssim 10$ \%) are characterized by a very low energy output. This happens when the BH
carries most of the collisional energy. In these cases, the nucleon remnant does not shower
and the visible energy is highly reduced by the increased graviton emission from the BH.
\section{Conclusions\label{concl}}
This paper focused on two aspects of BH airshowers which had been neglected in previous
studies: QCD and BH spin effects. The inclusion of these effects is important to check the
stability of the airshower profiles and provide a more accurate template for observational
searches. Event simulations based on the GROKE Monte Carlo show no change in the overall
characteristics of the airshowers. Effects due to color conservation, initial- and
final-state radiation, and different fragmentation models are washed out during the airshower
development. BH spin effects in the airshower development are estimated to be typically small
for two reasons: (i) most BHs are formed with low angular momentum and (ii) most of the CM
energy is carried by the nucleon remnant. It should be stressed, however, that a conclusive
statement on this issue requires the knowledge of the exact spinning BH greybody factors for
all fields in higher-dimensions.
\section*{Acknowledgements}
The authors thank Vitor Cardoso for many useful discussions and suggestions. This work was
supported (in part) by U.S.\ DoE contract DE-FG05-91ER40622. 

\end{document}